# A New Approach to Line – Sphere and Line – Quadrics Intersection Detection and Computation

Vaclav Skala

*Department of Computer Science and Engineering, Faculty of Applied Sciences, University of West Bohemia, Univerzitni 8, CZ 306 14 Plzen, Czech Republic*

**Abstract.** Line intersection with convex and un-convex polygons or polyhedron algorithms are well known as line clipping algorithms and very often used in computer graphics. Rendering of geometrical problems often leads to ray tracing techniques, when an intersection of many lines with spheres or quadrics is a critical issue due to ray-tracing algorithm complexity.

A new formulation of detection and computation of the intersection of line (ray) with a quadric surface is presented, which separates geometric properties of the line and quadrics that enables pre-computation. The presented approach is especially convenient for implementation with SSE instructions or on GPU.

**Keywords:** Line-sphere intersection, line-quadrics intersection, ray-tracing, implicit functions, tensor multiplication, projective space, homogeneous coordinates, clipping, GPU.
**PACS:** 02.60.-x , 02.30.Jr , 02.60 Dc

## INTRODUCTION

Spheres, ellipsoids and quadric surfaces are often used in geometric applications and in computer graphics as basic objects to build up more complex objects and scenes. Some techniques in computer graphics relay on bounding volumes, like spheres or ellipsoids, to significantly speed up computation, e.g. the intersection of a line (ray) with an object in ray tracing based methods [1], [2], [4]. Ray tracing is of $O(NM\,p\,2^k)$ computational complexity, where $N \times M$ is the final image resolution (typically $10^3 \times 10^3$), $p$ is a number of objects in a scene (typically $10^4 - 10^6$) and $k$ is a number of reflections/refractions of the light ray (typically $k = 5, ... ,8$). Considering a typical scene, we can easily reach $10^3[N].10^3[M].10^5[p].10^2[2^k] = 10^{13}$ of intersection tests. The complexity is even higher for time-varying scenes and stereoscopic rendering [5], [7], [13] etc.

Projective representation using homogeneous coordinates, i.e. a point $X = (X, Y, Z)$ is expressed in the projective space as $\boldsymbol{x} = [x, y, z{:}w]^T$, where $w \neq 0$ is the homogeneous coordinate and $x = w.X$, $y = w.Y$, $z = w.Z$, is often used not only in computer graphics, but "invisibly" in description of geometric objects defined in an implicit form, e.g. a plane $aX + bY + cZ + d = 0$ can be expressed in projective space as $ax + by + cy + dw = 0$ is defined by a vector $\boldsymbol{\rho} = [a, b, c{:}d]^T$. It is clearly visible, that $a, b, c$ are values without physical units, while $d$ has physical meaning, e.g. $[m]$. Now, the equation for a plane is given as $\boldsymbol{\rho}^T \boldsymbol{x} = 0$. It can be seen that the formula is more "compact" and convenient for matrix-vector computational architectures.

As the meaning of symbols $\boldsymbol{\rho}$ and $\boldsymbol{x}$ is indistinguishable, we have an example of a linear duality, which can be used in deriving new theorems and algorithms [8]-[9].

## LINE-QUADRICS INTERSECTION

Intersection of a line (ray) with a sphere, ellipsoid or with a general quadric surface is often used. Any quadric surface is given in implicit form as:

$$\boldsymbol{x}^T \boldsymbol{Q} \boldsymbol{x} = a_{11}x^2 + a_{22}y^2 + a_{33}z^2 + 2a_{12}xy + 2a_{13}xz + 2a_{23}yz + 2a_{14}x + 2a_{24}y + 2a_{34}z + a_{44} = 0 \quad (1)$$

where: $\boldsymbol{x} = [x, y, z{:}1]^T$ and the matrix $\boldsymbol{Q}$ is of the size $4 \times 4$ and symmetric.

The intersection of a line with a quadric surface is given by equations:

$$\boldsymbol{x}^T \boldsymbol{Q} \boldsymbol{x} = 0 \qquad \boldsymbol{x}(t) = \boldsymbol{x}_A + \boldsymbol{s}\, t \quad (2)$$

where: $\boldsymbol{s} = [s_x, s_y, s_z{:}0]^T$ and $\boldsymbol{x}_A = [x_A, y_A, z_A{:}1]^T$ in the Euclidean space or $\boldsymbol{s} = [s_x, s_y, s_z{:}s_w]^T$ and $\boldsymbol{x}_A = [x_A, y_A, z_A{:}w_A]^T$ in the projective space. Solving those equations we get a quadratic equation:

$$s^T Q s\, t^2 + 2 s^T Q x_A t + x_A^T Q x_A = 0 \quad (3)$$

the intersections of the line and the given quadratic surface is determined as:

$$at^2 + 2bt + c = 0 \qquad t_{1,2} = \frac{-b \pm \sqrt{b^2 - ac}}{a} \quad (4)$$

where: $a = s^T Q s$, $b = s^T Q x_A$ and $c = x_A^T Q x_A$.

If the discriminant $D = b^2 - ac < 0$, then there is no intersection of the line and the quadratic surface. As the intersection detection is especially important in computer graphics. Let us explore it more in detail.

## INTERSECTION DETECTION AND TENSOR MULTIPLICATION

In the discriminant computation, properties of the line and geometric properties of the quadratic surface are "mixed" together. The question is, whether they can be split and partially pre-computed to obtain faster computation.

From the Eq.(3), we can write:

$$D = (s^T Q x_A)(s^T Q x_A) - (s^T Q s)(x_A^T Q x_A) = s^T Q^T [x_A s^T - s x_A^T] Q x_A \\ = s^T Q^T [x_A \otimes s - s \otimes x_A] Q x_A = s^T Q^T R\, Q x_A \quad (5)$$

where: $Q$ matrix is a symmetric, $R$ matrix is an anti-symmetric matrix with a null diagonal and $\otimes$ stands for the tensor product, i.e. $a \otimes b = a b^T$.

$$R = \begin{bmatrix} 0 & x_A s_y - y_A s_x & x_A s_z - z_A s_x & -s_x \\ -(x_A s_y - y_A s_x) & 0 & y_A s_z - z_A s_y & -s_y \\ -(x_A s_z - z_A s_x) & -(y_A s_z - z_A s_y) & 0 & -s_z \\ s_x & s_y & s_z & 0 \end{bmatrix} \quad (6)$$

It can be seen that the $R$ matrix represents line (ray) properties. If the line is given by two points and $\sigma = x_B - x_A$ then:

$$R = \begin{bmatrix} 0 & x_A y_B - x_B y_A & x_A z_B - x_B z_A & x_A - x_B \\ -(x_A y_B - x_B y_A) & 0 & y_A z_B - y_B z_A & y_A - y_B \\ -(x_A z_B - x_B z_A) & -(y_A z_B - y_B z_A) & 0 & z_A - z_B \\ -(x_A - x_B) & -(y_A - y_B) & -(z_A - z_B) & 0 \end{bmatrix} = \begin{bmatrix} B & -\sigma \\ -\sigma^T & 0 \end{bmatrix} \quad (7)$$

The cross product $w \times v$ can be express in the matrix form as:

$$w \times v = \begin{bmatrix} 0 & -w_z & w_y \\ w_z & 0 & -w_x \\ -w_y & w_x & 0 \end{bmatrix} v \quad (8)$$

Therefore the $B$ matrix can be expressed by the matrix form of the cross product as:

$$Bq = \begin{bmatrix} 0 & x_A s_y - y_A s_x & x_A s_z - z_A s_x \\ -(x_A s_y - y_A s_x) & 0 & y_A s_z - z_A s_y \\ -(x_A s_z - z_A s_x) & -(y_A s_z - z_A s_y) & 0 \end{bmatrix} q = [\sigma \times \xi_A] \times q \quad (9)$$

where: $s = [s_x, s_y, s_z : 0]^T = [\sigma^T : 0]^T$ and $x_A = [x_A, y_A, z_A : 1]^T = [\xi_A^T : 1]^T$.

It means that elements of the $B$ matrix are given by a cross product $\sigma \times \xi_A$. The Eq. (5) can be rewritten as:

$$D = s^T Q^T [x_A \otimes s - s \otimes x_A] Q x_A = s^T Q^T R\, Q\, x_A \quad (10)$$

The above given formulation is easily applicable for cases, when a line (ray) is given by two points in the projective space as $x_A = [x_A, y_A, z_A : w_A]^T$ and $x_B = [x_B, y_B, z_B : w_B]^T$, i.e. linear interpolation with a monotonic non-linear parameterization is used [8], [11]. Then the $R$ matrix is given as:

$$R = \begin{bmatrix} 0 & x_A s_y - y_A s_x & x_A s_z - z_A s_x & x_A s_w - s_x w_A \\ -(x_A s_y - y_A s_x) & 0 & y_A s_z - z_A s_y & y_A s_w - s_y w_A \\ -(x_A s_z - z_A s_x) & -(y_A s_z - z_A s_y) & 0 & z_A s_w - s_z w_A \\ -(x_A s_w - s_x w_A) & -(y_A s_w - s_y w_A) & -(z_A s_w - s_z w_A) & 0 \end{bmatrix} \quad (11)$$

If the $M$ matrix is given as:

$$M = \begin{bmatrix} x_A & y_A & z_A & w_A \\ x_B & y_B & z_B & w_B \end{bmatrix} \quad (12)$$

then elements of the $R$ matrix are given as sub-determinants $M_{ij} = det[m_{i*}| m_{j*}]$ of the $M$ matrix, where: $m_{i*}$, resp. $m_{j*}$ is the $i$-th, resp. $j$-th column of the $M$ matrix. The $R$ matrix is given as:

$$R = \begin{bmatrix} 0 & M_{12} & M_{13} & M_{14} \\ -M_{12} & 0 & M_{22} & M_{24} \\ -M_{13} & -M_{13} & 0 & M_{34} \\ -M_{14} & -M_{24} & -M_{34} & 0 \end{bmatrix} \quad (13)$$

It should be noted that the $Q$ matrix defining a quadric surface is determined by the $Q_0$ matrix of the quadric surface in the fundamental position with application of a translation-rotational transformation given by the $T$ matrix, i.e.:

$$Q = T^T Q_0 T \quad (14)$$

As a sphere is the most used bounding volume in ray-tracing, let us explore it more in detail.

## LINE-SPHERE INTERSECTION DETECTION

As the geometrical scenes are defined by means of the Computer Solid Geometry (CSG), geometric object descriptions are kept in fundamental positions and lines (rays) are transformed by the inverse transform and the intersection detection is made. In the case of the intersection computation, the intersection point is transformed back accordingly. It means that some geometric transformations are made anyway. Therefore it is reasonable at accumulate transformation including the translation-rotational transformation given by the $T$ matrix together, i.e. a sphere the $Q_0$ matrix is defined as $Q_0 = diag[1,1,1,-r^2]$.

In this case the intersection detection of a line with a unit sphere in the fundamental position is significantly simplified. Let us assume that $s = [s_x, s_y, s_z: 0]^T$ and $x_A = [x_A, y_A, z_A: 1]^T$. The $R$ matrix *is constant* for the given line (ray), i.e.

$$R = [x_A \otimes s - s \otimes x_A] \quad (15)$$

As the sphere is in its fundamental position, we need to transform only the $x_A$ point of the line, as the *sphere rotationally invariant* and vector $s$ is "movable". Therefore the $D$ discriminant for the intersection detection is given as:

$$D = s^T Q^T R Q T^{-1} x_A$$

$$= [s_x, s_y, s_z: 0] \begin{bmatrix} 1 & 0 & 0 & 0 \\ 0 & 1 & 0 & 0 \\ 0 & 0 & 1 & 0 \\ 0 & 0 & 0 & -r^2 \end{bmatrix} [\ R\ ] \begin{bmatrix} 1 & 0 & 0 & 0 \\ 0 & 1 & 0 & 0 \\ 0 & 0 & 1 & 0 \\ 0 & 0 & 0 & -r^2 \end{bmatrix} \begin{bmatrix} 1 & 0 & 0 & -x_s \\ 0 & 1 & 0 & -y_s \\ 0 & 0 & 1 & -z_s \\ 0 & 0 & 0 & 1 \end{bmatrix} \begin{bmatrix} x_A \\ y_A \\ z_A \\ 1 \end{bmatrix}$$

$$= [s_x, s_y, s_z: 0] R \begin{bmatrix} 1 & 0 & 0 & 0 \\ 0 & 1 & 0 & 0 \\ 0 & 0 & 1 & 0 \\ 0 & 0 & 0 & -r^2 \end{bmatrix} \begin{bmatrix} x_A - x_s \\ y_A - y_s \\ z_A - z_s \\ 1 \end{bmatrix} = [s_x, s_y, s_z: 0] R \begin{bmatrix} x_A - x_s \\ y_A - y_s \\ z_A - z_s \\ -r^2 \end{bmatrix} \quad (16)$$

It can be seen that if a line is tested against many objects then detection of intersection is quite simple.
As the $R$ matrix is defined as:

$$R = \begin{bmatrix} B & -\sigma \\ \sigma^T & 0 \end{bmatrix} \quad (17)$$

and $\sigma = [s_x, s_y, s_z]^T$, $\delta_A = [x_A - x_s, y_A - y_s, z_A - z_s]^T$, we can write using Eq.(9):

$$D = [\sigma^T: 0] \begin{bmatrix} B & -\sigma \\ \sigma^T & 0 \end{bmatrix} \begin{bmatrix} \delta_A \\ -r^2 \end{bmatrix} = [\sigma^T B: -\sigma^T \sigma] \begin{bmatrix} \delta_A \\ -r^2 \end{bmatrix} = \sigma^T B \delta_A + r^2 \sigma^T \sigma$$
$$= \sigma^T [(\sigma \times \xi_A) \times \delta_A] + r^2 \sigma^T \sigma = \sigma^T \{[(\sigma \times \xi_A) \times \delta_A] + r^2 \sigma\} \quad (18)$$

It is necessary to point out, that:
- the expression $\sigma^T B$, resp. $(\sigma \times \xi_A)$ is constant for the given line (ray)
- modification for the projective space is simple and *no division operation is needed* as we can avoid transformation from the projective space to the Euclidean space, i.e. 6 division operations per a line are saved:

$$D = [\sigma^T: s_w] \begin{bmatrix} B & -\sigma \\ -\sigma^T & 0 \end{bmatrix} \begin{bmatrix} \delta_A \\ -r^2 \end{bmatrix} = [\sigma^T B - s_w \sigma^T: -\sigma^T \sigma] \begin{bmatrix} \delta_A \\ -r^2 \end{bmatrix} = \sigma^T B \delta_A - s_w \sigma^T \delta_A + r^2 \sigma^T \sigma \quad (19)$$

## CONCLUSION

This paper describes a new formulation of a line intersection computation with a sphere and with a quadric surface in $E^3$, in general. It separates geometric property of a line and quadric surfaces. This approach is especially convenient for cases when many lines intersections with a quadric surface are to be detected or/and computed, e.g. in ray tracing based methods of rendering geometric scenes. Due to the "matrix-vector" formulation, the proposed method is convenient for implementation with SSE instructions or GPU applications. It should be noted that the Eq.(16) can be simplified in the case of the specific quadric surface.

## ACKNOWLEDGMENTS

The author thanks to students and colleagues at the University of West Bohemia for recommendations, constructive discussions and hints that helped to finish the work. Many thanks belong to the anonymous reviewers for their valuable comments and suggestions that improved this paper significantly. This research was supported by the Ministry of Education of the Czech Republic – projects No. LH12181 and LG13047.

## APPENDIX

| | | | |
|---|---|---|---|
| Sphere | $x^2 + y^2 + z^2 - r^2 = 0$ $$\boldsymbol{Q}_0 = \begin{bmatrix} 1 & 0 & 0 & 0 \\ 0 & 1 & 0 & 0 \\ 0 & 0 & 1 & 0 \\ 0 & 0 & 0 & -r^2 \end{bmatrix}$$ | Ellipsoid | $\frac{x^2}{a^2} + \frac{y^2}{b^2} + \frac{z^2}{c^2} - 1 = 0$ $$\boldsymbol{Q}_0 = \begin{bmatrix} 1/a^2 & 0 & 0 & 0 \\ 0 & 1/b^2 & 0 & 0 \\ 0 & 0 & 1/c^2 & 0 \\ 0 & 0 & 0 & -1 \end{bmatrix}$$ |
| One-sheet Hyperboloid | $\frac{x^2}{a^2} + \frac{y^2}{b^2} - \frac{z^2}{c^2} - 1 = 0$ $$\boldsymbol{Q}_0 = \begin{bmatrix} 1/a^2 & 0 & 0 & 0 \\ 0 & 1/b^2 & 0 & 0 \\ 0 & 0 & -1/c^2 & 0 \\ 0 & 0 & 0 & -1 \end{bmatrix}$$ | *Hyperbolic paraboloid | $\frac{x^2}{a^2} - \frac{y^2}{b^2} - 2z = 0$ $$\boldsymbol{Q}_0 = \begin{bmatrix} 1/a^2 & 0 & 0 & 0 \\ 0 & -1/b^2 & 0 & 0 \\ 0 & 0 & 0 & -1 \\ 0 & 0 & -1 & 0 \end{bmatrix}$$ |

\* Notice: The $\boldsymbol{Q}_0$ matrix is not diagonal
Similarly for other quadric surfaces.